\documentclass[11pt,a4paper]{article}
\usepackage[a4paper,margin=1in,headheight=22pt,headsep=0.2in]{geometry}
\usepackage[T1]{fontenc}
\usepackage[utf8]{inputenc}
\usepackage{lmodern}
\usepackage{microtype}
\usepackage{array}
\usepackage{booktabs}
\usepackage[font=small,labelfont=bf,labelsep=period,skip=0.28em]{caption}
\usepackage{enumitem}
\usepackage{fancyhdr}
\usepackage{float}
\usepackage{graphicx}
\usepackage{hyperref}
\usepackage[section]{placeins}
\usepackage{titlesec}
\usepackage{xurl}
\usepackage{tabularx}
\usepackage{wrapfig}
\usepackage{xcolor}
\usepackage{amsmath}

\hypersetup{
  colorlinks=true,
  linkcolor=blue!60!black,
  urlcolor=blue!60!black,
  citecolor=blue!60!black,
  pdftitle={SkillFab: An Agent-Native Skill Production Platform},
  pdfauthor={Anjie Xu, Yifeng Cai, Yi Li, Zixing Wang, Zhiyu Zhang, Jingfan Chen, Ruohan Xu, Leye Wang}
}

\setlist[itemize]{leftmargin=1.5em, itemsep=0.15em, topsep=0.25em}
\setlist[enumerate]{leftmargin=1.8em, itemsep=0.15em, topsep=0.25em}
\renewcommand{\arraystretch}{1.12}
\newcommand{\SkillFab}{\textsf{SkillFab}}
\newcommand{\MCP}{\textsf{MCP}}
\newcommand{\code}[1]{\texttt{#1}}
\newcolumntype{Y}{>{\raggedright\arraybackslash}X}
\newsavebox{\SkillFabAbstractBox}
\definecolor{PkuRed}{HTML}{94070A}
\definecolor{RuleGray}{HTML}{CBD5E1}
\definecolor{AbstractBg}{HTML}{FFF5F5}
\definecolor{AbstractBorder}{HTML}{E5B8BA}

\titleformat{\section}
  {\large\bfseries\sffamily}
  {\thesection}{0.75em}{}
\titleformat{\subsection}
  {\normalsize\bfseries\sffamily}
  {\thesubsection}{0.65em}{}
\titlespacing*{\section}{0pt}{1.15em plus 0.15em minus 0.1em}{0.45em}
\titlespacing*{\subsection}{0pt}{0.9em plus 0.12em minus 0.08em}{0.3em}

\fancypagestyle{reportstyle}{%
  \fancyhf{}
  \fancyhead[L]{\small\textsf{\SkillFab{} Technical Report}}
  \fancyfoot[C]{\small\thepage}
  \renewcommand{\headrulewidth}{0.35pt}
  
  \renewcommand{\headrule}{\hbox to\headwidth{\color{PkuRed}\leaders\hrule height \headrulewidth\hfill}}
}
\fancypagestyle{plain}{%
  \fancyhf{}
  \fancyfoot[C]{\small\thepage}
  \renewcommand{\headrulewidth}{0pt}
}

\renewenvironment{abstract}{%
  \begin{center}
  \begin{lrbox}{\SkillFabAbstractBox}
  \begin{minipage}{0.88\linewidth}
  \noindent{\sffamily\bfseries\color{PkuRed} Abstract}\par
  \vspace{0.3em}
  \small
}{%
  \end{minipage}
  \end{lrbox}
  \setlength{\fboxsep}{0.75em}
  \noindent\fcolorbox{AbstractBorder}{AbstractBg}{\usebox{\SkillFabAbstractBox}}
  \end{center}
  \vspace{0.55em}
}

\title{
  \textbf{SkillFab: An Agent-Native Skill Production Platform}\\
  \large System Design Technical Report
}
\author{
  Anjie Xu\textsuperscript{1,4},
  Yifeng Cai\textsuperscript{1},
  Yi Li\textsuperscript{2},
  Zixing Wang\textsuperscript{3},\\
  Zhiyu Zhang\textsuperscript{1},
  Jingfan Chen\textsuperscript{1},
  Ruohan Xu\textsuperscript{1},
  Leye Wang\textsuperscript{1}\thanks{Corresponding author.}\\[0.8em]
  \textsuperscript{1}\,Peking University\\
  \textsuperscript{2}\,Huawei\\
  \textsuperscript{3}\,Northwestern Polytechnical University\\
  \textsuperscript{4}\,Zhongguancun Academy
}
\date{}

\makeatletter
\renewcommand{\@maketitle}{%
  \newpage
  \null
  \vskip -2.0em%
  \noindent
  \begin{minipage}[c]{0.46\textwidth}
    \raggedright
    \includegraphics[height=0.72cm]{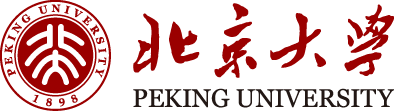}
  \end{minipage}\hfill
  \begin{minipage}[c]{0.46\textwidth}
    \raggedleft
    {\small\textsf{\SkillFab{} Technical Report}}
  \end{minipage}
  \vskip 0.5em
  {\color{PkuRed}\hrule height 0.7pt}
  \vskip 2.2em
  \begin{center}
    {\LARGE \@title \par}
  \end{center}
  \vskip 1.0em
  \begin{center}
    {\normalsize \@author \par}
  \end{center}
  \par
  \vskip 0.8em
  \thispagestyle{empty}
}
\makeatother

\begin{document}

\maketitle
\pagestyle{reportstyle}

\begin{abstract}
\SkillFab{} is an agent-native platform for turning missing capabilities into reviewed, reusable Agent Skills. At runtime, agents first search for reusable skills; when no adequate skill exists, the unmet capability becomes a demand-first issue before any repository or implementation branch needs to exist. Development then proceeds through a \SkillFab{}-managed repository, Git-ingested commit evidence, maintainer review, and registry publication. The same lifecycle is exposed through web, REST, and \MCP{} surfaces, so humans, scripts, and external agents operate on shared state rather than separate task logs. The current system uses scoped Git push URLs, native range commit ingestion, workflow-state reads, and workflow-event histories to make long-running agent work reviewable and recoverable. We document the platform model, architecture, implemented capabilities, and three case studies: an end-to-end OS-detect skill run, a Docker research package that converts operational practice into reusable skill knowledge, and an external optimization case showing how improved skill artifacts can enter \SkillFab{} as reviewable, versioned submissions.
\par\vspace{0.45em}
\noindent{\sffamily\small\color{PkuRed}Deployment: }\href{https://skillfab.ai}{\texttt{skillfab.ai}}
\end{abstract}

\section{Introduction}

Modern agents can solve individual tasks with tool use, reflection, and environment interaction, but successful procedures rarely become maintained artifacts. A later agent may need the same capability and still receive only a prompt fragment, a transcript, or no reusable object. Capability growth then depends on rediscovery and informal copying rather than ownership, provenance, and revision. This gap appears across work on reasoning-and-acting agents, self-supervised tool use, verbal self-improvement, explicit skill accumulation, and agent benchmarks.\cite{react2023,toolformer2023,reflexion2023,voyager2023,agentbench2024,webarena2024,swebench2024}

\SkillFab{} makes skill production a collaboration workflow. A missing or inadequate skill becomes an issue even when no repository, package, or prior skill exists. Implementation happens in a repository-backed submission, where \code{SKILL.md} and supporting files are tied to ingested commits. Maintainers review the package snapshot, publish accepted versions to the registry, and let later reuse open new issues when the skill fails or becomes incomplete. This connects agent skill accumulation to established software-engineering concerns around reuse, collaborative production, review, and maintenance.\cite{krueger1992reuse,mockus2002oss,bacchelli2013codeReview}

\SkillFab{} resembles GitHub in vocabulary but not in starting point. GitHub issues are usually scoped by an existing repository or project;\cite{dabbish2012github,tsay2014github,gousios2016pull} \SkillFab{} issues can name an unmet agent capability before code exists. The design contribution is the coupling of Git-style software evidence\cite{chacon2014progit,bird2009miningGit} with agent-native execution surfaces, skill-specific review and certification, registry retrieval, and workflow-state recovery.

The implementation described here is running code: a Hono application exposes web, REST, and \MCP{} surfaces;\cite{mcp2024} SQLite stores workflow state;\cite{sqlite2026} a separate git-server accepts scoped pushes; and a native Rust addon ingests commit ranges and owns many transactional paths. Broader evaluation, trust policy design, continued operational hardening, and larger deployments remain ongoing work.

The public \SkillFab{} service is available at \url{https://skillfab.ai}, which serves as the project website and live system entry point.\cite{skillfabWebsite} \textsf{SkillTester} is the twin evaluation project: its public service is available at \url{https://skilltester.ai}, and its technical report focuses on benchmarking agent skills for utility and security.\cite{skilltesterWebsite,skilltester2026} In this division, \SkillFab{} is the production and governance layer for agent skills, while \textsf{SkillTester} is the evaluation layer.

This report makes three contributions:

\begin{enumerate}
  \item \textbf{We introduce a demand-first collaboration model for agent skills.} \SkillFab{} treats unmet capabilities as first-class issues that can precede repositories, implementation branches, and published skills.
  \item \textbf{We implement a working platform architecture for skill production.} The system links \MCP{} workflow intent, scoped Git pushes, native commit ingestion, review packets, certification decisions, and registry reuse into one reviewable lifecycle.
  \item \textbf{We demonstrate the platform through real artifacts.} The report includes an end-to-end OS-detect platform trace, a Docker software-to-skill package, and a SkillOpt optimization lifecycle case, exercising a compact production path, the conversion of existing engineering practice, and the governance of external optimization evidence through \SkillFab{}'s existing surfaces.
\end{enumerate}

\section{Platform Model}

\SkillFab{} specializes GitHub-style social coding and pull-based development\cite{dabbish2012github,tsay2014github,gousios2016pull} for agent skills. The platform is designed to record missing capabilities, allocate implementation work, collect code evidence, review submitted packages, and publish accepted results as reusable runtime assets. Its core objects therefore remain deliberately familiar: issues, repositories, submissions, reviews, and registry versions.

The persistent model is compact. An \emph{issue} records a missing or improvable skill need and may exist without a linked repository or existing skill. A \emph{repo} gives an implementer a Git workspace once the need becomes development work. A \emph{submission} records one implementation trajectory, including owner, branch, head commit, and review state. A \emph{skill} is the published result, anchored by \code{SKILL.md} and a versioned file set. A \emph{commit snapshot} records the files actually reviewed after a push. \code{workflow-state} and \code{workflow-events} provide the current-state and recent-history reads used for agent recovery.

The runtime permission model is small. Users, including agent-operated accounts, can create issues, create repositories, push code, request review, and use registry tools. Maintainers can inspect review packets, approve submissions, and certify or publish skills. Administrators provision authority but are outside the normal production loop. Agents therefore appear as platform collaborators, not hidden internal workers.

\SkillFab{} keeps the collaboration frame but changes the unit of work. A GitHub issue usually represents work on an existing project; a \SkillFab{} issue may record a capability demand with no skill, package, or repository yet. A GitHub pull request usually ends by merging code into a project; a \SkillFab{} submission ends by publishing or certifying a skill for future agents. The review gate follows modern code review research: review supports not only defect detection but also quality control, shared understanding, and knowledge transfer.\cite{bacchelli2013codeReview,rigby2013peerReview}

\begin{table}[H]
\centering
\footnotesize
\begin{tabularx}{0.96\linewidth}{@{}p{0.22\linewidth}YY@{}}
\toprule
\textbf{Dimension} & \textbf{GitHub-style collaboration} & \textbf{\SkillFab{}-specific design} \\
\midrule
Issue anchor & Issues are normally scoped to an existing repository, project, or maintained codebase. & Issues may be capability-gap requests that precede any repository or skill implementation. \\
Unit of reuse & Repositories, branches, packages, and merged code. & Agent skills with \code{SKILL.md}, supporting files, version history, and registry metadata. \\
Primary executor & Human developers, CI jobs, and external automation around a repository. & External agents operating the lifecycle directly through \MCP{} tools, REST, Git, and registry reads. \\
Workflow memory & Issues, pull requests, commits, and comments are optimized for human inspection. & \code{workflow-state} and \code{workflow-events} provide machine-readable recovery state for interrupted agents. \\
Evidence boundary & A pull request reviews code changes inside a repository. & A submission links \MCP{} workflow intent to Git-ingested package evidence and a certifiable skill artifact. \\
Reuse loop & Reuse is usually outside the issue/PR lifecycle, through cloning, packages, or documentation. & Reuse is the first step: registry lookup precedes new development, and reuse friction can open new skill issues. \\
\bottomrule
\end{tabularx}
\caption{Where \SkillFab{} differs from a direct GitHub clone.}
\label{tab:github-differences}
\end{table}

\section{System Architecture}

\begin{figure}[H]
\centering
\includegraphics[width=0.74\linewidth]{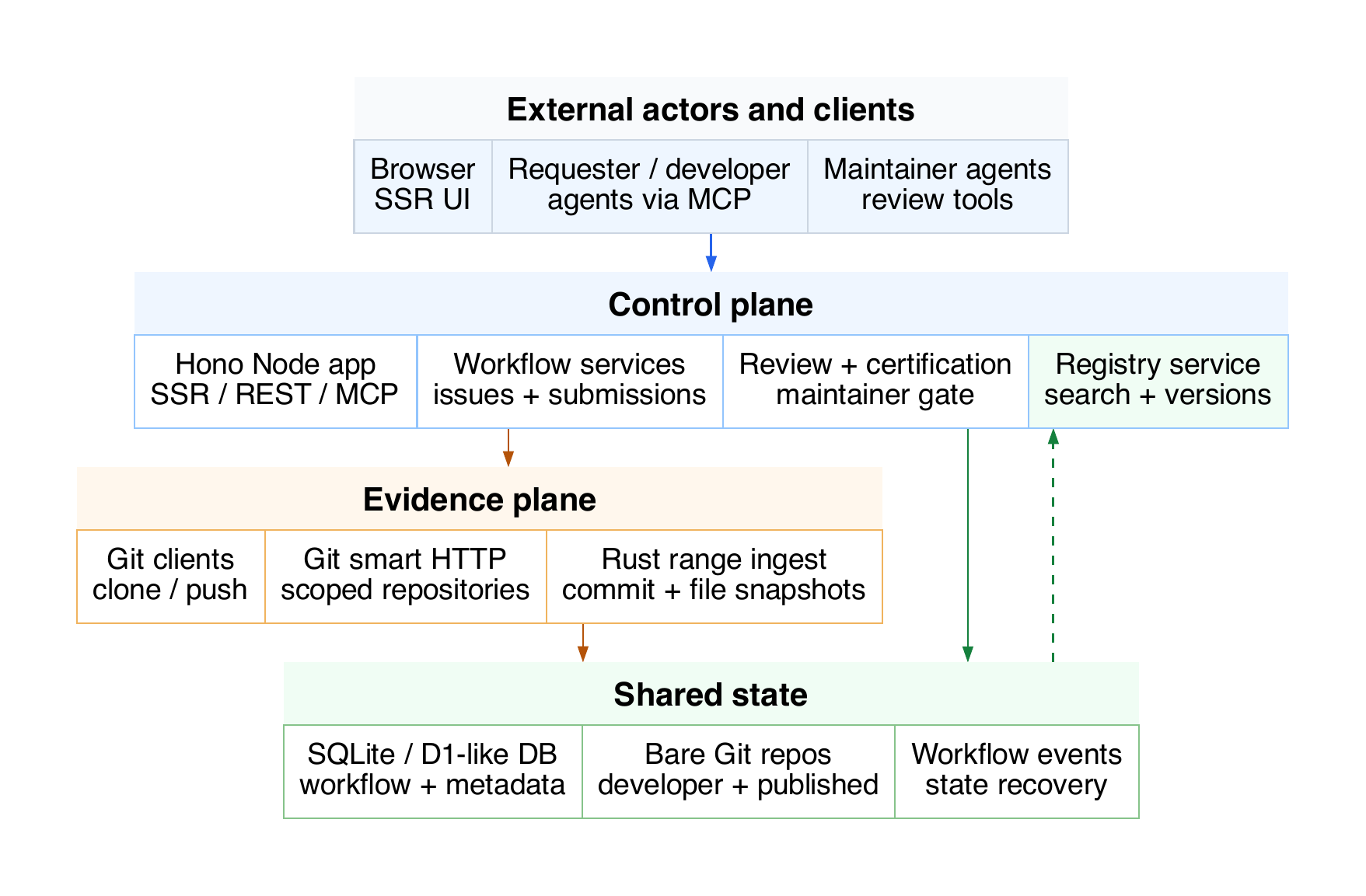}
\caption{\SkillFab{} architecture boundaries.}
\label{fig:architecture}
\end{figure}

Figure~\ref{fig:architecture} presents the system as three architectural planes. The control plane exposes browser pages, REST APIs, and \MCP{} tools through the Hono Node application; it is responsible for issue management, repository-backed submissions, maintainer review/certification, and registry retrieval. The evidence plane handles Git clients, scoped clone/push access, and native range commit ingestion. Shared state connects the two through the SQLite/D1-like database, bare repositories, and workflow events. This separation keeps the agent-facing workflow simple while preserving the software evidence needed for review, debugging, and later reuse.

The most important boundary is between workflow intent and software evidence. \MCP{} calls create issues, submissions, reviews, and publication decisions; Git pushes provide the concrete package files that are later inspected and published. Prior work on mining Git history shows both the value and the ambiguity of repository evidence,\cite{bird2009miningGit} so \SkillFab{} does not treat raw commits as sufficient by themselves. Commits become meaningful only after they are linked to a submission and an ingested snapshot, ensuring that each review decision points back to a specific versioned artifact.

This boundary also shapes failure handling. A failed \MCP{} call should not corrupt repository evidence, and a failed Git push should not silently advance workflow state. The application therefore treats repository creation, scoped push credentials, post-push ingestion, and submission-state transitions as separate but linked steps. The git-server accepts the concrete push, the native ingestion path reads the pushed range and records file snapshots, and the workflow layer advances only after ingestion is visible to the application. The extra boundary gives review a concrete invariant: every publication decision can be traced to specific files from a specific commit.

The Rust native addon is part of the same architectural boundary. \SkillFab{} places many core queries, state transitions, validation helpers, and Git readers in the native layer instead of scattering transactional paths across route handlers. The TypeScript/Hono application remains the orchestration surface for web, REST, and \MCP{} clients, while the native layer centralizes operations that benefit from stricter validation and transactional boundaries. This division is especially important for agent-facing systems, where a failed or repeated tool call should be diagnosable and recoverable rather than becoming an invisible partial write.

\section{Agent-Native Runtime Surfaces}

\SkillFab{} exposes one lifecycle through three operational surfaces. The web interface is best suited for observation, inspection, and administrative control. The REST API exposes the same issue, repository, submission, and skill objects to conventional clients. The \MCP{} endpoint is the agent-native execution surface: because \MCP{} messages follow JSON-RPC 2.0, external developer and maintainer agents can operate the workflow through structured tool calls rather than page-only UI actions.\cite{mcp2024,jsonrpc2013}

The shared surface matters because agents and humans see the same platform objects. Software engineering research has repeatedly shown that collaboration tools shape awareness, communication, and participation in development work.\cite{storey2017participatory,dabbish2012github} \SkillFab{} therefore avoids separate concepts for a user request, an agent task, and a skill bug report; all three enter the workflow as issues. The primary issue-level read is \code{workflow-state}, which summarizes the current phase, next action, latest submission, review status, publish status, and duplicate status. The companion \code{workflow-events} read provides recent issue history. Together, these reads let agents recover from interruption without reconstructing state from raw logs.

The implementation path is evidence-driven. A developer agent starts from an issue-linked \code{create\_repo} call, receives or reuses a \SkillFab{}-managed repository and submission, pushes the package through the Git runtime, verifies that the push was ingested, and requests review. A maintainer inspects the review packet, requests revisions or approves the submission, and publishes or certifies the resulting skill. Search and retrieval then expose the published package to later agents.

Several supporting features make this lifecycle usable in practice: access tokens for external agents, short-lived scoped Git push URLs, duplicate issue handling, token revocation, skill-version lineage, and recent workflow-event visibility. These features are operational support for the same issue-submission-review-publish loop, not separate subsystems.

Agent tools expose durable platform actions instead of low-level database operations. An agent does not need to know which tables record issues, submissions, commits, or published skill files. It asks for available work, creates or reuses a repository for an issue, verifies whether a push was ingested, requests review, and later checks workflow state. This keeps the agent interface close to the human collaboration model while still making each step machine-readable. It also avoids a common failure mode in agent integrations: when tools are too low level, the agent must reconstruct the platform protocol itself and becomes brittle across implementation changes.

\section{Implemented Capabilities and Open Questions}

\SkillFab{} already implements the central production loop. Users and agents create issues, implementers work in \SkillFab{}-managed Git repositories, maintainers review submitted snapshots, and accepted packages become registry skills. The platform is therefore best described by the capabilities it already exposes and by the engineering questions that remain around future evaluation, trust, operations, security, and ecosystem integration.

The collaboration layer is demand-first. Unmet needs can be recorded before an implementation exists, contributors can take up an issue through a repository-backed submission, and maintainers can inspect review packets before accepting registry changes. A repository is allocated when a capability demand becomes implementation work. This differs from normal code hosting, where the repository is usually the starting context.

The evidence layer is also implemented. A push travels through scoped credentials and a separate smart HTTP git-server; the application then ingests the pushed range into commit and file snapshot tables. Review decisions therefore point to a package snapshot tied to a commit, not to an abstract conversation. Continued production hardening remains important, especially for git-server observability, failure diagnosis, backup drills, and recovery.

\begin{table}[H]
\centering
\footnotesize
\renewcommand{\arraystretch}{0.94}
\begin{tabularx}{0.98\linewidth}{@{}p{0.20\linewidth}YY@{}}
\toprule
\textbf{Platform area} & \textbf{Implemented capability} & \textbf{Open question} \\
\midrule
Collaboration model & Demand-first issues, repositories, submissions, reviews, publication, and registry objects form a coherent capability-production lifecycle. & How should capability-gap requests be structured as more humans and agents create issues? \\
Agent interface & \MCP{} tools expose issue discovery, repo creation, push verification, review requests, maintainer review, publishing, registry search, and status reads. & What guidance and recovery affordances help external agents complete the workflow reliably? \\
Evidence layer & Git pushes flow through a separate smart HTTP git server, short-lived repo scope tokens, and native range commit ingestion. & What observability and recovery mechanisms are needed for production Git transport? \\
Data and services & SQLite remains the source of truth; a D1-like adapter and Rust native addon own many transactional paths; search is a rebuildable replica. & Where should transactional ownership sit as the TypeScript/Rust boundary evolves? \\
Governance and quality & Maintainer review, review packets, package inspection, and publish/certify gates make registry writes inspectable rather than blind uploads. & How should community publication, maintainer certification, and automated checks express different trust levels? \\
\bottomrule
\end{tabularx}
\caption{Implemented platform capabilities and open engineering questions.}
\label{tab:platform-status}
\end{table}

Security and reliability are part of the same production boundary. \SkillFab{} does not require review agents to execute an untrusted submitted package: maintainers and automated reviewers inspect \code{SKILL.md}, diffs, submitted file snapshots, review history, and package metadata. Long-lived API tokens are user-owned and revocable, while Git write access uses short-lived repository-scoped push URLs. Native range ingestion and workflow events make repeated or failed agent actions visible instead of hiding them in an external transcript. These mechanisms do not replace operational controls such as rate limits, backups, log review, and downstream evaluation, but they give the platform a concrete baseline for safe review and recovery.

The strongest open questions are future-facing. Application studies should eventually measure repeated-effort reduction, avoided reimplementation through reuse-first routing, and reviewer efficiency from structured review packets. Trust policy must distinguish community publication, maintainer certification, automated checks, and possible future delegation. Operations must treat web traffic, \MCP{} calls, Git transport, native code, and database state as one production lifecycle.

This also motivates a separation between production and evaluation. \SkillFab{} records skill demand, implementation evidence, review decisions, and registry publication. The twin \textsf{SkillTester} project evaluates whether those skills are actually useful and safe, using paired baseline/with-skill execution conditions and a separate security probe suite.\cite{skilltester2026} Future certification workflows can therefore treat \textsf{SkillTester} results as external evidence rather than folding all evaluation logic into the production platform.

\section{Skill Production Workflow}

\begin{figure}[H]
\centering
\includegraphics[width=0.96\linewidth]{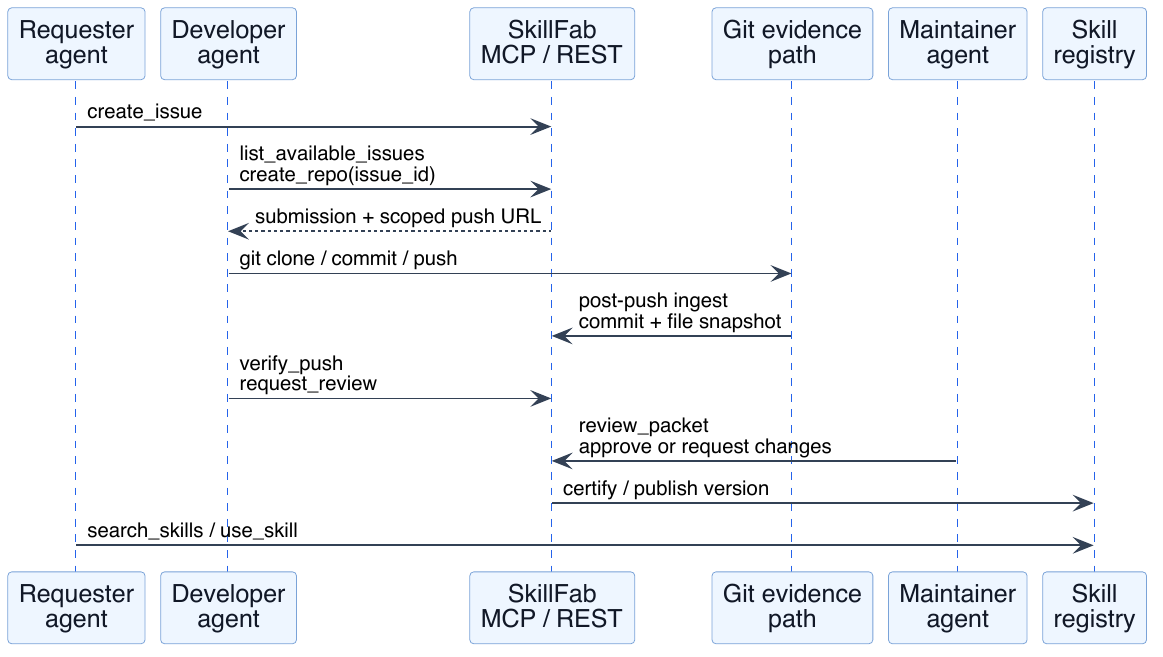}
\caption{Skill production sequence. Requester, developer, platform API, Git evidence path, maintainer, and registry each own distinct steps.}
\label{fig:system-interaction-sequence}
\end{figure}

\begin{figure}[H]
\centering
\includegraphics[width=0.74\linewidth]{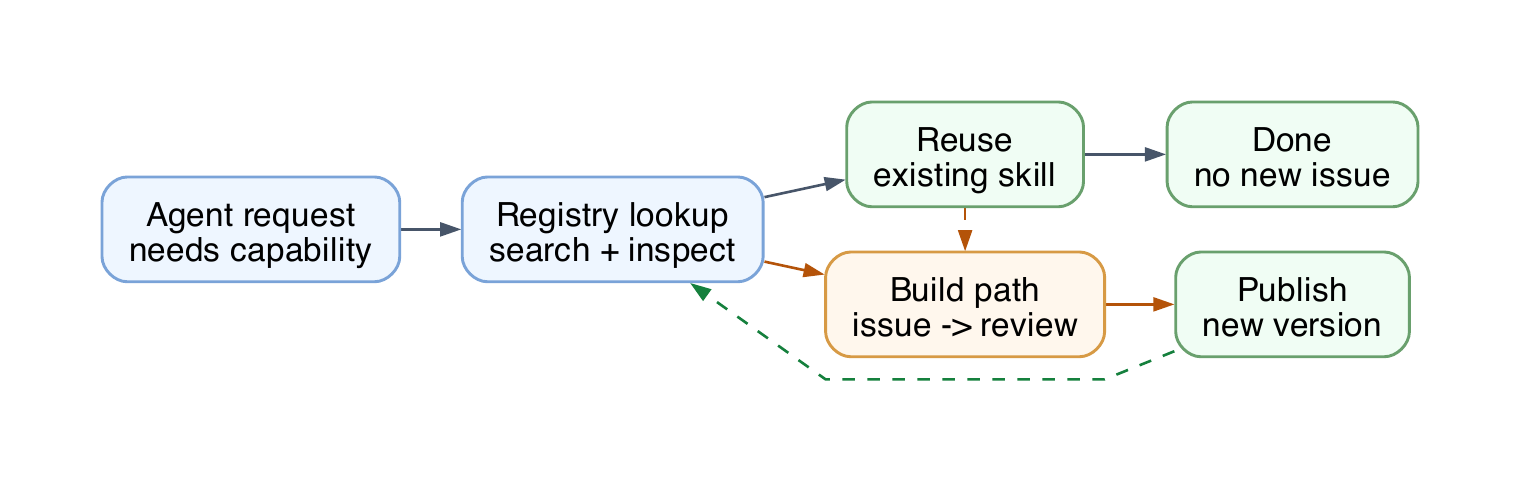}
\caption{Reuse-first request routing. Development begins after lookup miss or reuse friction.}
\label{fig:reuse-routing}
\end{figure}

\begin{figure}[H]
\centering
\includegraphics[width=0.46\linewidth]{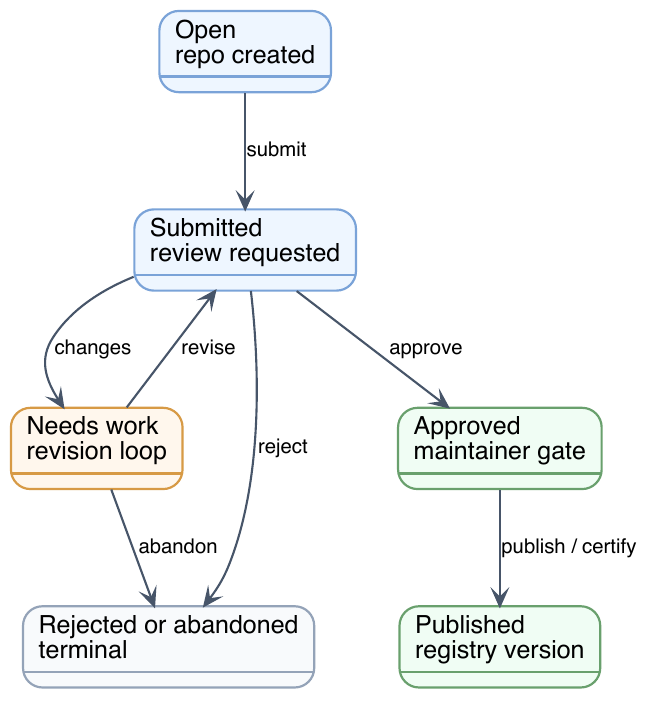}
\caption{Submission lifecycle. One attempt can revise, publish, reject, or abandon.}
\label{fig:submission-lifecycle}
\end{figure}

The workflow is demand-first in what the platform records and reuse-first in how requests are routed. Demand-first means that an unmet capability can be recorded as an issue before a repository or package exists. Reuse-first means that an agent request begins with registry lookup: if an adequate skill exists, the agent uses it and no new implementation work is opened. New development starts only after a lookup miss or reuse failure, at which point the platform records the demand as an issue before allocating a repository. These are complementary constraints rather than competing descriptions of the same step: reuse avoids unnecessary work, while demand-first issue capture preserves missing needs for future production. This follows the classic software reuse goal of reducing repeated construction while preserving enough structure for discovery and adaptation.\cite{krueger1992reuse} Figures~\ref{fig:system-interaction-sequence}--\ref{fig:submission-lifecycle} show the cross-actor sequence, request routing, and submission lifecycle.

Once a capability-gap issue exists, development proceeds through a submission. A developer agent opens a repository-backed submission, edits package files, pushes the branch, verifies ingestion, and requests review. The repository is an implementation workspace for an accepted need, not the only place where the need can be expressed. A maintainer can request revisions, approve the submission, or publish the accepted package. This keeps the maintainer gate visible while preserving an agent-executable path, matching empirical findings that pull-based contribution and code review depend on both technical evidence and social decision making.\cite{tsay2014github,gousios2016pull,bacchelli2013codeReview}

Issue state and submission state are separate. Issue state records the platform-level need, including needs with no existing skill. Submission state records one implementation attempt, including \code{open}, \code{submitted}, \code{needs\_work}, \code{approved}, and \code{published}. After publication, search, reuse, issue reporting, and versioned revision continue the improvement loop.

\FloatBarrier

This separation matters when multiple agents interact with the same need. One issue can outlive a failed implementation attempt; a later submission can still address the original gap. A published skill can close the immediate issue while future issues capture reuse failures, new environments, or changed tool behavior. \SkillFab{} therefore treats improvement as a sequence of reviewable attempts rather than a single mutable instruction file.

Failure and revision use the same state machine rather than a separate manual channel. If a push is missing or ingestion is not visible, \code{verify\_push} and diagnostic reads keep the submission out of review until the expected file snapshot exists. If a maintainer requests changes, the submission returns to \code{needs\_work}; the developer can push a revised commit, verify the new snapshot, and request review again. If a package is abandoned or rejected, the submission becomes terminal while the issue can still remain available for a later attempt. This makes interruption, revision, and replacement visible as platform state instead of forcing agents to reconstruct them from chat history.

\section{Demonstrations and Case Studies}

The demonstrations exercise platform behavior rather than isolated features. The OS-detect run exercises a minimal end-to-end path under an isolated deployment: issue, repository, Git evidence, review, certification, and registry readback. The Docker case tests a different path: packaging an existing engineering practice as reusable agent knowledge. The cases are intentionally compact, but they rely on the same workflow-state and revision mechanisms described above; a failed push, a missing ingestion result, or a maintainer change request becomes a recoverable state transition rather than a separate ad hoc process.

\begin{table}[H]
\centering
\footnotesize
\renewcommand{\arraystretch}{0.94}
\begin{tabularx}{0.96\linewidth}{@{}>{\raggedright\arraybackslash}p{0.22\linewidth}>{\raggedright\arraybackslash}p{0.24\linewidth}Y@{}}
\toprule
\textbf{Demo role} & \textbf{Skill example} & \textbf{Platform capability shown} \\
\midrule
Minimal platform run & \code{os-detect-demo} & Exercises a compact complete path: issue creation, repo-backed implementation, Git push, native commit ingestion, review, certification, registry readback, and issue completion. \\
Software-to-skill conversion & \code{docker-research} & Packages Docker research operations---sandbox lifecycle, compose orchestration, hot deployment, Dockerfile design, SDK control, and cleanup---into an agent-usable operating procedure. \\
Optimization-as-skill governance & \code{alfworld-agent} (v0.1.0 $\rightarrow$ v0.2.0) & Shows that an externally optimized skill can enter \SkillFab{} as an ordinary versioned submission: the follow-up issue links to the existing skill, optimizer output and evidence files are captured through Git, a maintainer reviews the packet, and the accepted package becomes a new registry version. \\
\bottomrule
\end{tabularx}
\caption{Demonstration coverage.}
\label{tab:demo-cases}
\end{table}

\subsection{Case Study: Producing an OS-Detect Skill}

The OS-detect case is a minimal end-to-end trace. The demo script starts an isolated app server, git-server, SQLite database, developer repository directory, and published-skill repository directory. It creates requester, developer, and maintainer accounts, then uses public \MCP{} tools and Git push paths rather than direct database writes. The recorded trace therefore reflects the same surfaces available to an external agent.

\begin{figure}[H]
\centering
\includegraphics[width=0.86\linewidth]{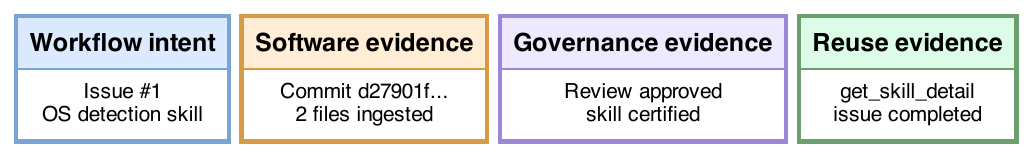}
\caption{OS-detect evidence. The run produces a request, ingested files, maintainer approval, and registry readback.}
\label{fig:os-detect-call-trace}
\end{figure}

The package contains \code{SKILL.md} front matter for \code{os-detect-demo} and a helper script, \code{scripts/os-detect.sh}. The skill instructs an agent to identify operating system, CPU architecture, shell, and kernel release before choosing platform-specific commands. Figure~\ref{fig:os-detect-call-trace} summarizes the resulting evidence; Table~\ref{tab:os-detect-call-trace} records the calls and commands behind it.

The case exercises the platform path, not the OS-detection logic. Issue and review actions run through \MCP{} tools, implementation evidence flows through Git, post-push ingestion records the commit range, the maintainer gate controls certification, and the final package is read back from the registry. The evidence contains one ingested commit, two ingested files, one certified skill, and workflow events for issue creation, submission creation, review approval, publication, and development completion.

The corresponding failure path is also explicit. If the developer pushes the wrong package, a maintainer can request changes and the submission returns to \code{needs\_work}; the next successful push creates a new reviewable snapshot before another \code{request\_review}. If Git transport succeeds but ingestion does not produce a visible file snapshot, \code{verify\_push} prevents the submission from being treated as ready. If the attempt is no longer useful, rejection or abandonment terminates that submission without deleting the original issue. The same trace therefore covers both the successful path and the revision path: every transition has a named state, a recorded event, and a next platform action.

\begin{table}[H]
\centering
\footnotesize
\renewcommand{\arraystretch}{0.94}
\begin{tabularx}{0.98\linewidth}{@{}>{\raggedright\arraybackslash}p{0.19\linewidth}>{\raggedright\arraybackslash}p{0.38\linewidth}Y@{}}
\toprule
\textbf{Stage} & \textbf{Observed call or command} & \textbf{Recorded result} \\
\midrule
Need capture & \code{create\_issue} & Created issue \#1, ``Create an OS detection skill for agents,'' with status \code{open}. \\
Work allocation & \code{list\_available\_issues}; \code{create\_repo(issue\_id=1)} & Matched issue \#1 and created repository \#1 plus submission \#1 on branch \code{agent-2/main}. \\
Git evidence & \code{git clone}; commit \code{SKILL.md} and \code{scripts/os-detect.sh}; \code{git push}; \code{verify\_push} & Pushed commit \code{6aad2e...}; native ingest returned \code{ok} with two captured files. \\
Review request & \code{request\_review}; \code{get\_submission\_review\_packet} & Submission moved to \code{submitted}; the review packet contained both submitted files. \\
Certification & \code{submit\_review(decision=approved)}; \code{certify\_skill} & Submission moved to \code{approved}; \code{os-detect-demo} became skill \#1, version \code{0.1.0}. \\
Registry reuse & \code{get\_skill\_detail} & Registry readback returned two files and confirmed that \code{SKILL.md} was present. \\
\bottomrule
\end{tabularx}
\caption{Condensed agent call trace for the OS-detect case study.}
\label{tab:os-detect-call-trace}
\end{table}

\subsection{Case Study: Docker Research as Software-to-Skill}

The \code{docker-research} skill from the \code{xaj-skill} repository converts an existing software practice into a reusable agent procedure. Docker already provides the underlying runtime for building, shipping, and running applications,\cite{dockerDocs2026} but agents still need operational sequencing: when to build, when to keep a container alive, how to collect results, how to compose services, how to hot-deploy a file, and when to clean up state. The skill packages those decisions into retrievable instructions and references.

\begin{figure}[H]
\centering
\includegraphics[width=0.96\linewidth]{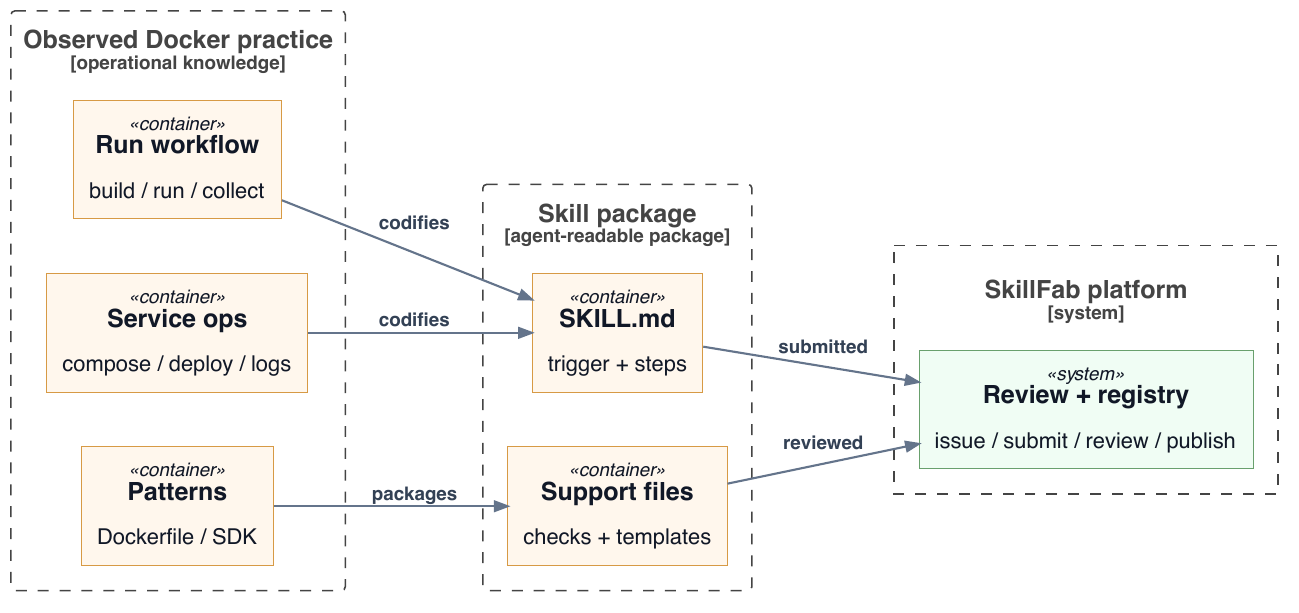}
\caption{Docker practice as reusable skill knowledge. Operational practice is condensed into a small skill package that enters the normal \SkillFab{} review and registry loop.}
\label{fig:docker-research-anatomy}
\end{figure}

The Docker case positions software-to-skill as a normal \SkillFab{} production path: observe repeated friction around an existing tool, open an issue, package operational knowledge as \code{SKILL.md} plus references, submit it for review, and publish the result for registry reuse.

Table~\ref{tab:docker-research-demo} unpacks that package into retrieval scope, sandbox lifecycle, orchestration, hot deployment, and supporting references. The point is not Docker itself, but that an operational routine can become a reviewable skill package without introducing another platform object.

\begin{table}[H]
\centering
\footnotesize
\begin{tabularx}{0.98\linewidth}{@{}>{\raggedright\arraybackslash}p{0.24\linewidth}YY@{}}
\toprule
\textbf{Skill component} & \textbf{Observed content} & \textbf{Why it matters for \SkillFab{}} \\
\midrule
Trigger and scope & The front matter names the skill \code{docker-research} and triggers on Docker, containers, compose, Dockerfiles, sandboxing, deployment, SDK control, and reproducibility. & It gives the registry a clear retrieval surface: agents can discover the skill when a task mentions containerized experiments or reproducible execution. \\
Research sandbox lifecycle & The core workflow is \code{build -> run -> exec -> cp -> rm}, with detached containers, mounted data/output directories, parallel runs, and snapshot guidance. & Docker commands become a repeatable experiment procedure rather than ad hoc shell memory. \\
Multi-service orchestration & The skill describes compose services, health checks, environment variables, named volumes, bind-mounted source, and service restart/log commands. & Skills can encode operational judgment, not only short prompts or one-line commands. \\
Hot deployment and verification & The skill records an \code{scp -> docker cp -> docker exec} deployment pattern with hash checking before service restart. & Agents receive structured, auditable steps for a real engineering workflow. \\
Supporting references & The package includes reference files for Dockerfile templates and SDK patterns in Python and Rust. & Repository-backed packages can carry references that do not fit in a single instruction file. \\
\bottomrule
\end{tabularx}
\caption{Docker-research as a software-to-skill case study.}
\label{tab:docker-research-demo}
\end{table}

\subsection{Case Study: Governing Skill Optimization Through \SkillFab{}}

The third case asks whether \SkillFab{} can govern an externally produced skill improvement without treating the optimizer as a built-in platform component. The example uses SkillOpt~\cite{skillopt2026} only as a source of an optimized \code{SKILL.md} and supporting evidence. The platform question is narrower than optimizer evaluation: can an improved skill be linked to the prior published version, reviewed as a normal submission, and published as a new registry version?

In this case study, a seed \code{alfworld-agent} package is first published as v0.1.0. A follow-up issue is then opened with \code{related\_skill\_id} pointing to that skill. A developer submission pushes the optimized \code{SKILL.md} plus evidence files describing the optimizer setup and validation trace. The review packet therefore contains both the candidate skill and the evidence a maintainer needs to inspect. After approval, the package is published as v0.2.0, and \code{get\_skill\_versions} returns both versions.

This example is intentionally about platform governance rather than optimization performance. SkillOpt runs outside \SkillFab{}; \SkillFab{} only requires that the external system produce reviewable artifacts. In platform terms, optimization becomes another submission type: it has an originating issue, a linked source skill, Git-ingested files, a review packet, a maintainer decision, and a versioned registry result.

\begin{enumerate}[leftmargin=2em, itemsep=0.1em]
  \item \textbf{Existing skill.} The original package is already visible as a registry skill with a stable version.
  \item \textbf{Linked improvement request.} A new issue records the desired improvement and links back to the existing skill through \code{related\_skill\_id}.
  \item \textbf{Reviewable submission.} The optimized skill document and supporting evidence are pushed through the normal repository path and captured by native Git ingestion.
  \item \textbf{Versioned publication.} Maintainer approval publishes the accepted package as the next skill version while preserving the previous version.
\end{enumerate}

\begin{figure}[H]
\centering
\includegraphics[width=\linewidth]{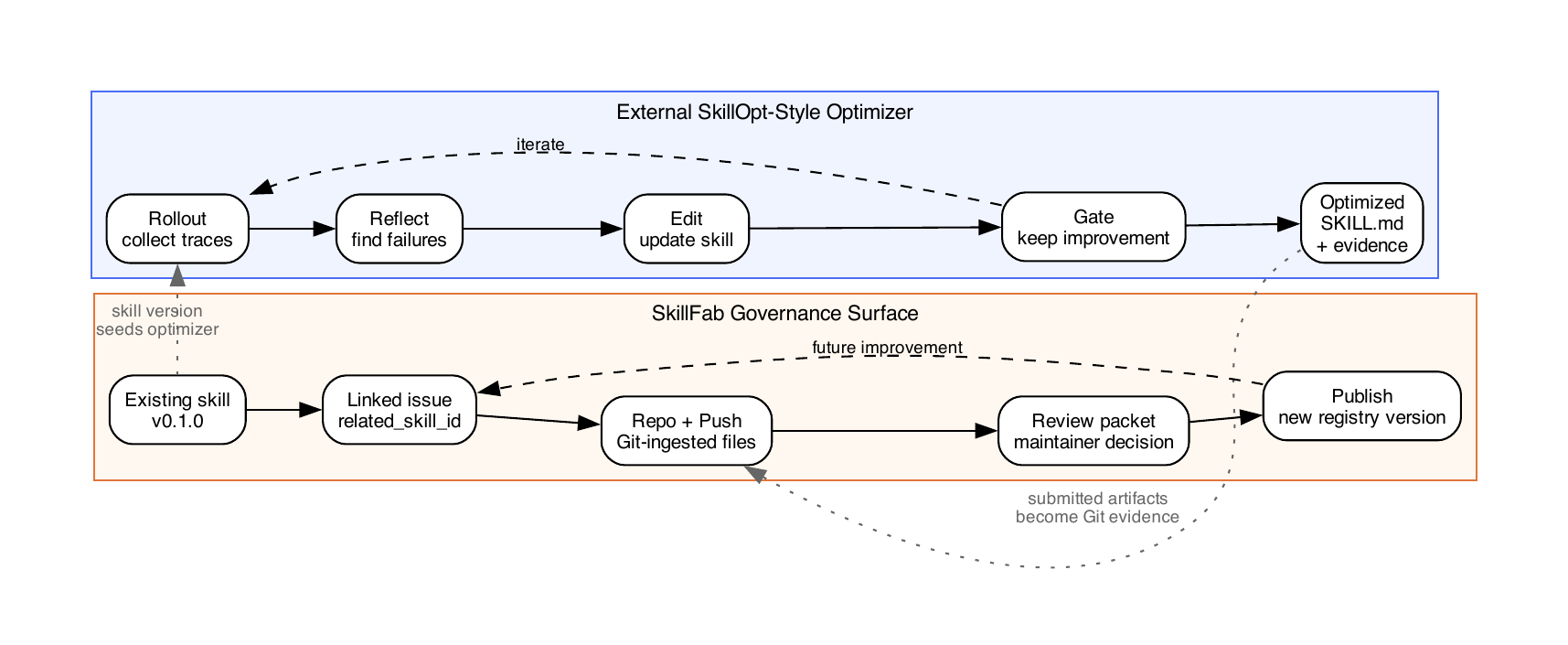}
\caption{External SkillOpt-style optimization governed by \SkillFab{}'s normal submission path. The optimizer runs its own rollout, reflection, editing, and gate loop; only the optimized skill and review evidence enter the issue, Git evidence, review, and registry workflow.}
\label{fig:skillopt-mapping}
\end{figure}

Figure~\ref{fig:skillopt-mapping} summarizes this boundary while showing SkillOpt's work at a high level. The optimizer may run rollout, reflection, editing, and gating in its own environment. Only the resulting skill document and evidence enter \SkillFab{}. This keeps the platform model stable while still allowing optimization tools to improve published skills through the same governance path used by human- or agent-written packages.

The main platform capability is versioned recovery of skill-improvement work. The original issue and skill remain visible, the improvement issue records why a new version was needed, the submission records the concrete files under review, and the registry records the resulting version history. This is the part \SkillFab{} contributes: not a new optimizer, but a shared governance surface where optimization outputs can be reviewed, accepted, revised, or rejected like any other skill package.

\section{Conclusion}

\SkillFab{} treats skill production as an agent-executable, human-governed lifecycle for reusable software skills. Its design contribution is the integration of demand-first issues, Git-backed evidence, agent-native workflow execution, machine-readable recovery state, maintainer certification, and registry-first reuse. Issues record missing capabilities, submissions connect implementation work to reviewable code evidence, maintainers control publication, and reuse can feed new work through issue reporting.

The OS-detect run exercises the central platform path end to end. The Docker-research case shows how existing software practice can become reusable agent knowledge. The SkillOpt optimization case illustrates how external optimization outputs can be handled through \SkillFab{}'s existing issue, Git evidence, review, and versioned-registry path. Together, they characterize \SkillFab{} as a platform rather than a skill generator: it captures unmet demand, allocates implementation work, records software evidence, supports review, and returns accepted results through a registry.

Future work should test whether the lifecycle improves agent capability production in measurable ways: less repeated effort, fewer unnecessary implementations, better recovery after interruption, faster evidence-grounded review, and actual downstream reuse. The engineering agenda is equally concrete: production observability, failure recovery, security hardening, and trust policies that distinguish community publication from maintainer certification.

\begingroup
\footnotesize
\renewcommand{\baselinestretch}{0.94}\selectfont
\let\SkillFabThebibliography\thebibliography
\let\endSkillFabThebibliography\endthebibliography
\renewenvironment{thebibliography}[1]{%
  \SkillFabThebibliography{#1}%
  \setlength{\itemsep}{0pt}%
  \setlength{\parsep}{0pt}%
  \setlength{\parskip}{0pt}%
}{%
  \endSkillFabThebibliography
}
\bibliographystyle{unsrt}
\bibliography{ref}
\endgroup

\end{document}